\documentclass{article}

\usepackage{arxiv}

\usepackage[utf8]{inputenc} 
\usepackage[T1]{fontenc}    
\usepackage{hyperref}       
\usepackage{amsthm} 

\usepackage{url}            
\usepackage{booktabs}       
\usepackage{amsfonts}       
\usepackage{nicefrac}       
\usepackage{microtype}      
\usepackage{lipsum}
\usepackage{graphicx}
\graphicspath{ {./images/} }

\usepackage{mathtools}
\usepackage{amssymb}

\usepackage{algorithm}    
\usepackage{algorithmic}  

\usepackage[capitalize,nameinlink,noabbrev]{cleveref} 

\crefname{theorem}{Theorem}{Theorems}
\crefname{definition}{Definition}{Definitions}
\crefname{proposition}{Proposition}{Propositions}
\crefname{lemma}{Lemma}{Lemmas}
\crefname{corollary}{Corollary}{Corollaries}

\makeatletter
\@addtoreset{equation}{section} 
\makeatother

\title{Generating Temporal Networks with the Ascona Model}

\author{
 Samuel Koovely \\
  Department of Mathematical Modeling and Machine Learning\\
  University of Zurich\\
  Zurich, Switzerland \\
  \texttt{samuel.koovely@math.uzh.ch} \\
}

\begin{document}
\maketitle
\begin{abstract}
  We introduce a queueing-based sampling framework for continuous-time temporal networks. We focus on a Markovian parametrization in which link start times follow a homogeneous Poisson process and link durations are exponentially distributed. We derive stochastic properties of the resulting link streams and exploit them to generate synthetic temporal networks with controllable smoothness and prescribed event patterns, relevant for the validation and interpretation of methods for community, scale, change-point, and periodicity detection. By coupling this temporal mechanism with block-structured endpoint distributions, we obtain a continuous-time analogue of stochastic block models. We also discuss extensions of the framework, including discrete-time and instantaneous-contact limits.
\end{abstract}


\section{Introduction}
Temporal (or dynamic) networks/graphs are extensions of graphs that incorporate temporal information. Various mathematical objects can be considered to be temporal networks, and different formalisms have been proposed to describe them (e.g., \cite{casteigts_time-varying_2012, holme_temporal_2012, latapy_stream_2018}). These objects can be classified on a spectrum of increasing complexity based on the granularity of the encoded temporal information \cite{rossetti_social_2015, rossetti_community_2018}. On one side, we have static graphs, which aggregate in one structure all temporal information; on the other side, there are stream graphs and link streams that avoid any aggregation. Somewhere in between the two sides lie snapshot networks and discrete-time stream graphs/link streams.

Temporal networks are relevant in modeling many complex systems \cite{masuda_guide_2016}, and, as for their static counterparts, understanding their communities/clusters is a fundamental problem in network science.
The community structure evolution is typically characterized by archetypal structural changes \cite{cazabet_challenges_2023}, called \emph{events}, or combinations thereof \cite{failla_describing_2024}. Therefore, sampling methods that allow modeling these archetypes are of paramount importance for the generation of quality synthetic data with controlled ground-truth. One needs these synthetic networks, for instance, for the validation and interpretation of algorithms tasked for community, scale, change-point, and periodicity detection. A review of the main existing methods, particularly relevant for community detection benchmarks, can be found in \cite{cazabet_evaluating_2021}.

For some applications, one would also want to sample networks that are less artificial and instead are closer to real-world datasets. Recently proposed techniques in this direction (e.g., \cite{gauvin_randomized_2022, longa_efficient_2022, presigny_building_2021}) aim at producing surrogate temporal graphs, structures that maintain some or most properties of a reference (often real-world) temporal network. However, these methods do not provide sufficient control to design a controlled experimental setup, as they bypass the explicit formulation of a modeling mechanism by relying on a pre-existing structure.

Agent-based models (e.g., \cite{perra_activity_2012, snijders_introduction_2010}) are an alternative option. They leverage control of microscopic dynamics of a complex system and combine them so that, in the aggregate, complex phenomena emerge. However, often, the understanding of these models relies heavily on simulations, as the underlying dynamics are complex, leading to mathematical intractability, hence limited stochastic and statistical control.

Most of the temporal network literature, especially on sampling methods, has focused on snapshot networks or discrete-time structures, leaving out the more complex continuous-time ones (a notable exception being  \cite{bovet_flow_2022}). The main contribution of this article is the introduction of a sampling method for continuous-time temporal networks, and the explanation of how to use it to sample data reflecting the above-mentioned event patterns, plus a continuous-time generalization of the stochastic block model (SBM) \cite{abbe_community_2018}.

The model works in two fundamental steps: in the first, we sample when nodes are connected with a link, and in the second, we sample which nodes to assign to each link sampled in the previous step. Separating the temporal from the connectivity aspect offers benefits in terms of implementation and clarity, while not sacrificing expressivity.
This separation mechanism also helps to structure the mathematical analysis of a temporal network (model), as the two aspects can be considered orthogonal to each other, hence treated separately, as in \cite{barucca_maximum_2025}. Connectivity can exhibit almost arbitrary levels of complexity and is responsible for various structural aspects such as assortativity, disassortativity, and nestedness. A complete analysis of its properties from a stochastic and statistical perspective is therefore challenging.
The temporal aspect appears to be simpler; its probabilistic and statistical analysis can be studied \cite{barucca_maximum_2025, sanna_passino_modelling_2019, sanna_passino_mutually_2023}. This article follows the footsteps of these works: its key idea is to describe the appearance and disappearance of nodes' connections in terms of queueing processes, and then focus on the evolution of the total number of active connections as a function of time to derive probabilistic and statistical results.

Queueing processes were originally introduced to describe the flow of customers (jobs, particles, calls) arriving at a service facility. It is possible to approximate the evolution of most queueing systems with Markov chains. The advantage of these is that it is possible to calculate the stationary distribution of a chain by solving a linear system of equations called the balance conditions \cite{walrand_introduction_1988}. \\

The structure of this work is as follows. In \cref{sec:Theory} we introduce the notation and describe the objects studied in the article. It is divided into two parts: the first is devoted to formalisms of temporal networks, the second to queueing processes. In \cref{sec:Ascona} we introduce the Ascona model, a temporal networks model based on queueing theory, and then use a Markovian parametrization of it to generate link streams with desired temporal evolutions.
Based on this Markovian framework, we describe a link stream generalization of SBMs, and compare it to dynamic SBMs, a popular temporal extension of them. We conclude the section with a description of variations of the Ascona model. The end of the article, \cref{sec:Conclusions}, contains a discussion of the presented results and highlights possible future research directions stemming from this work. 

\section{Theoretical Framework} \label{sec:Theory}
In this section, we describe the notation and objects we need to introduce the Ascona model. In the first part, we describe various formalisms of temporal networks following the framework presented in \cite{latapy_stream_2018}, and in the second part, general queueing processes, focusing on the $M/M/ \infty$ queue, also known as the ample-server model \cite[Ch. 2.7]{gross_fundamentals_2008}.

\subsection{Temporal Networks} \label{subsec:temp_nets}
We model a dynamic network within a \emph{time domain} $T$ (either an interval of $\mathbb{R}$ for continuous time or a finite subset of $\mathbb{Z}$ for discrete time) and over a finite \emph{node set} $V$. In the most general setup, stream graphs, $V$ is not fixed in time. However, for many, if not most applications, a disappearing node is equivalent to a node with zero degree that never disappears. Therefore, in \cref{sec:Ascona}, we focus on the goal of producing a sampling method for link streams, where $V$ is constant.
In \cref{sec:Conclusions}, we are going to briefly discuss how one could extend the model to generate stream graphs.

\subsubsection{Link Streams} \label{subsubsec:linkstreams}
The \emph{link stream} is a temporal network formalism that concentrates on the timing of interactions. There are two common variants.

\subsubsection*{General Model}
A link stream is a triplet
\begin{equation}
    L=(T, V, E), \qquad 
E\subseteq \bigl\{ [\alpha,\omega)\times\{\{u,v\}\} : \alpha \leqslant \omega, u\neq v \bigr\},
\end{equation}

so that $([\alpha,\omega),\{u,v\})\in E$ means the link $\{u,v\}$ is present for all $t\in[\alpha,\omega)$. \\

For each $t\in T$ the \emph{instantaneous graph} is
\begin{equation}
    G_t = (V, E_t), \quad
    E_t \coloneqq \bigl\{\{u,v\} : (t,\{u,v\})\in E \bigr\}.
\end{equation}

Given a window $I=[a,b]\subseteq T$, the \emph{unweighted footprint} is the static graph
\begin{equation} \label{eq:footprint_u}
    G_{I} \coloneqq \left(V, \bigl\{\{u,v\} : \exists t\in I \text{ with } (t,\{u,v\})\in E \bigr\}\right).
\end{equation}
If $T$ is not discrete, the \emph{weighted footprint} assigns to each edge the contact time
\begin{equation} \label{eq:footbprint_w}
    w_I(u,v) \coloneqq \lambda\left(\{t\in I : (t,\{u,v\})\in E\}\right),
\end{equation}
where $\lambda$ is the Lebesgue measure. If $T$ is discrete, we use the contact count instead.

We say that a stream $L'=(T', V', E')$ is a \emph{sub-stream} of $L=(T, V, E)$ if, intuitively, $T' \subset T, V' \subset V, E' \subset E$.

\subsubsection*{Instantaneous Contacts}
The second variant is a special case of a link stream where links have 0-duration. A link stream with instantaneous events is a triple
\begin{equation}
    L=(T, V, E), \qquad E \subseteq T \times \binom{V}{2},
\end{equation}
where $(t,\{u,v\})\in E$ records a contact of $\{u,v\}$ at time $t$.

\subsubsection{Snapshot Networks}
The most common framework to incorporate temporal information in networks consists of splitting the time domain into slices and then building a graph, called a snapshot, for each time slice: its nodes and links represent the interactions that occurred during this time slice.
Consequently, snapshot networks consist of a discrete set of graphs and are structurally similar to the above-mentioned discrete-time streams. 

The length of each slice is determined based on a compromise: one needs slices large enough to ensure that each snapshot captures significant structural information, but large slices lead to losses of temporal information, since all interactions within the same slice are merged. In addition, several or even varying slice durations may be relevant \cite{latapy_stream_2018}.

The above-described models are all natural generalizations of simple graphs to the temporal context and will be the objects of study of this article. One may enrich them by attaching attributes to links or nodes, without changing the underlying $(T,V,E)$ structure, and obtain natural extensions of multiedges, weighted, and signed networks.

\subsection{The $M/M/\infty$ Queue}
Queues were originally introduced to model the flow of customers arriving at a service facility. They are modeled stochastically as birth-death Markov chains keeping track of the number of elements in the system. The basic ingredients of a queue model are:
\begin{itemize}
  \item an arrival process $A$ (typically described by interarrival times),
  \item a service-time distribution $S$,
  \item a number $c\in\mathbb{N}\cup\{\infty\}$ of parallel servers,
  \item a system capacity $K$ that limits the number of customers  that can be present,
  \item $\bar{K}$ which is the maximal number of elements entering the system,
  \item a service discipline $D$ that specifies how customers are selected for service (e.g., first-come, first-served)
\end{itemize}
Kendall's notation, originating from \cite{kendall_stochastic_1953}, concisely encodes these features as
\begin{equation}
    A/S/c/K/\bar{K}/D.
\end{equation}
 Common defaults are $K=\infty$, $\bar{K}=\infty$, and $D=\text{first-come, first-served (FCFS)}$, in which case one writes simply \(A/S/c\). 

We are mainly interested in one specific queue, the $M/M/ \infty$ one. The `$M$' is a standard symbol in Kendall notation indicating Markovian processes; therefore, the model can be interpreted as a birth (or immigration)-death process where arrivals form a Poisson process with rate $\lambda>0$, and deaths are i.i.d. exponential with rate $\mu>0$ \cite{asmussen_applied_2003}. There are countably infinitely many servers, hence no waiting occurs: each arrival immediately starts the death process. \\
Let $M(t)$ denote the number of customers in the system at time $t\ge 0$. Then $\{M(t)\}_{t\ge 0}$ is a continuous–time birth–death process on $\mathbb{N}$ with birth rates
\begin{equation}
    q_{m,m+1}=\lambda \quad \text{for all } m\ge 0,
\end{equation}
and death rates
\begin{equation}
    q_{m,m-1}=m \mu \quad \text{for } m\ge 1,
\end{equation}
since each of the $m$ customers present departs at rate $\mu$ independently. \\
The transition rate matrix of the process is
\begin{equation}
    Q=\begin{pmatrix}
-\lambda & \lambda \\
\mu & -(\mu+\lambda) & \lambda \\
&2\mu & -(2\mu+\lambda) & \lambda \\
&&3\mu & -(3\mu+\lambda) & \ddots \\
&&& \ddots&\ddots
\end{pmatrix}.
\end{equation}
Writing $p_m(t)=\mathbb{P}(M(t)=m)$, the corresponding Kolmogorov forward equations are
\begin{align}
p_0'(t) &= -\lambda p_0(t) + \mu p_1(t),\\
p_m'(t) &= \lambda p_{m-1}(t) - (\lambda+m\mu)p_m(t) + (m+1)\mu p_{m+1}(t), \qquad m \geqslant 1.
\end{align}

\subsubsection{Process Evolution}
\begin{figure}[htbp]
  \centering
    \includegraphics[width=0.95\linewidth]{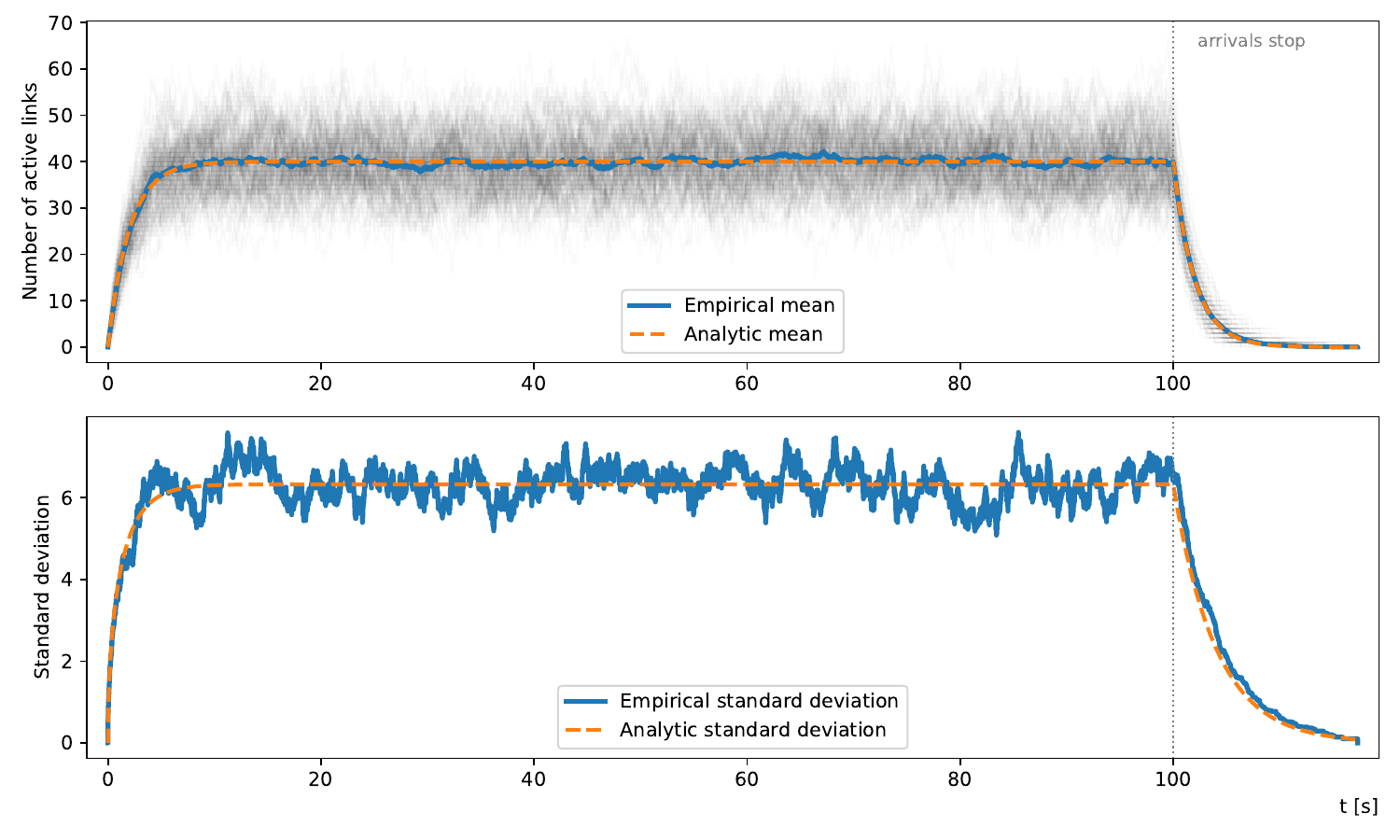}
  \caption{ Comparison of analytical formulas and trajectories of 100 $M/M/\infty$ queue samples. In the top plot, we compare the average density of the sample as a function of time with the analytical formula eq. \eqref{eq:mean_profile}. The fluctuations of the trajectories are in the square root order of the mean profile, as predicted analytically. The comparison of this analytical formula for the standard deviation with the sample one is displayed in the bottom plot.}
  \label{fig:trajectories_m_sd}
\end{figure}

It is known (e.g., \cite[Ch. 6]{robert_stochastic_2013}) that, assuming that $M(0)=0$,  $M(t)$ is Poisson with a time–dependent mean
\begin{equation} \label{eq:poisson_queue}
    m(t)=\frac{\lambda}{\mu}\left(1-e^{-\mu t}\right) \footnote{For a general initial condition
    $m(t)=m(0)e^{-\mu t}+\frac{\lambda}{\mu}\left(1-e^{-\mu t}\right)$};
\end{equation}
hence, for every $j\in\mathbb{N}$, the transition probabilities from $0$ are
\begin{equation}
\mathbb{P}\{M(t)=j\mid M(0)=0\}
=\exp\!\left(-\frac{\lambda}{\mu}\left(1-e^{-\mu t}\right)\right)
\frac{\left(\frac{\lambda}{\mu}(1-e^{-\mu t})\right)^j}{j!}.
\end{equation}
We can derive this fact by observing that the probability generating function $G(z,t) \coloneqq \sum_{m\ge 0}p_m(t) z^m$ solves 

\begin{equation}
    \frac{\partial G(z,t)}{\partial t}=(z-1)\lambda G(z,t) -(z-1)\mu \frac{\partial G(z,t)}{\partial z}
\end{equation} with $G(z,0)\equiv 1$; yielding
\begin{equation}
    G(z,t)=\exp\!\Big(m(t)(z-1)\Big),
\end{equation}
the probability generating function of a $\mathrm{Poisson}(m(t))$ law.
\\
This gives
\begin{equation}
    \mathbb{E}[M(t)]=\frac{\lambda}{\mu}\left(1-e^{-\mu t}\right),
\end{equation}
and
\begin{equation}
    \mathrm{Var}\!\left(M(t)\right)=\frac{\lambda}{\mu}\left(1-e^{-\mu t}\right).
\end{equation}
More generally, if we start the process at a starting time $t_s$, and stop sampling elements after a certain time $t_e$, then the process is shifted and truncated, and the mean follows the equation
\begin{equation}\label{eq:mean_profile}
    m(t)=
    \begin{cases}
        0, & t<t_{s},\\[4pt]
        \frac{\lambda}{\mu}\left(1-e^{-\mu(t-t_{s})}\right), & t_{s}\le t\le t_e,\\[4pt]
        \frac{\lambda}{\mu} \left(e^{-\mu(t-t_e)}-e^{-\mu(t-t_{s})}\right), & t>t_e.
    \end{cases}
\end{equation}
We display these properties in \cref{fig:trajectories_m_sd}.

\subsubsection{Stationary Regime}
The $M/M/ \infty$ chain is always irreducible and positive recurrent; this is in contrast with finite servers queues $M/M/c$, where $\frac{\lambda}{c \mu} < 1$ is required for ergodicity. So, if we do not stop sampling after a certain time-point $t_e$, we converge towards a stationary behavior.

One can get the asymptotic behavior by noticing that  $m(t)\to \lambda/\mu$ and that the Poisson family is closed under weak limits, hence
\begin{equation}
    M(t)\ \xrightarrow[]{d}\ \mathrm{Poisson}\!\left(\frac{\lambda}{\mu}\right).
\end{equation}
Alternatively, the stationary probabilities $\{\pi_m\}_{m\ge 0}$ satisfy the detailed balance equations
\begin{equation}
    \pi_m\lambda=\pi_{m+1}(m+1)\mu \quad \text{for } m\ge 0,
\end{equation}
expressing equilibrium between the birth and death processes of the queue.
Solving recursively gives
\begin{equation}
    \pi_m=\pi_0\frac{(\lambda/\mu)^m}{m!}.
\end{equation}
Normalizing with $\sum_{m=0}^\infty \pi_m=1$ yields
\begin{equation}
    \pi_m = \mathbb{P}\{M(\infty)=m\} = e^{-\rho}\frac{\rho^m}{m!}, \qquad m=0,1,2,\dots
\end{equation}
i.e., the stationary distribution is $\mathrm{Poisson}(\rho)$ with mean $\rho \coloneqq \frac{\lambda}{\mu}$. 
\\
Consequently,
\begin{equation}
    \mathbb{E}[M(\infty)]=\operatorname{Var}(M(\infty))=\frac{\lambda}{\mu}.
\end{equation}
One could also recover $\mathbb{E}[M(\infty)]= \frac{\lambda}{\mu}$
directly from Little's law (see \cite{little_proof_1961} and \cite[Thm. 3.4.1]{walrand_introduction_1988}).

We now derive an estimate for the time $t^\star$ at which we can assume that $m(t)$ is close enough to the stationary value. We use this estimate later in \cref{subsubsec:archetypes} and \cref{subsec:generate_snapshots}.

\subsubsection{Head of the Queue} \label{subsec:head_queue}

We consider an $M/M/\infty$ queue that starts empty at time $t=0$, with arrival rate $\lambda>0$ (Poisson arrivals) and i.i.d.\ service times
$S \sim \mathrm{Exp}(\mu)$ of mean $\frac{1}{\mu}>0$.
\\
One can heuristically think of being at steady state as if the system had ``forgotten'' its empty start, so the mean occupancy is close to
\begin{equation}
m_\infty \coloneqq \frac{\lambda}{\mu}.
\end{equation}
We define the (deterministic) distance from stationarity at time $t$ as
\begin{equation}
\delta(t)
\coloneqq
m_\infty - m(t)
=
\frac{\lambda}{\mu}  e^{-\mu t}.
\label{eq:delta_def}
\end{equation}
Because $M(t)$ is Poisson with mean $m(t)$, its standard deviation at time $t$
is
\begin{equation}
\sigma(t)
=
\sqrt{\mathrm{Var}[M(t)]}
=
\sqrt{m(t)}
=
\sqrt{\frac{\lambda}{\mu} \left(1 - e^{-\mu t}\right)}.
\label{eq:sigma_def}
\end{equation}
We now formalize the idea that the system has become ``indistinguishable
from stationarity'' as the first time $t^\star$ when the initial
bias $\delta(t)$ is no larger than the typical intrinsic fluctuation scale
$\sigma(t)$.
Concretely, we are interested in the crossover time $t^\star$ defined by
\begin{equation}\label{eq:criterion}
\delta(t^\star) = \sigma(t^\star).
\end{equation}
The computations in \cref{sec:appendix_head} show that
\begin{equation} \label{eq:tstar_exact}
t^\star
=
-\frac{1}{\mu}
\ln\!\left(
\frac{-1 + \sqrt{1 + 4 \frac{\lambda}{\mu}}}{2 \frac{\lambda}{\mu}}
\right),
\end{equation}
and that if $\frac{\lambda}{\mu}$ is moderately large, we can approximate it as
\begin{equation} \label{eq:tstar_approx}
    t^\star \approx \frac{1}{2 \mu} \ln(\frac{\lambda}{\mu}).
\end{equation}

In the case of $N$ queue samples, the standard deviation of the sample mean is $\sqrt{\frac{m(t)}{N}}$, so it takes more time to establish closeness to stationarity. The computation is equivalent to that of the single sample case. This heuristic argument can be made rigorous in terms of the total variation distance, but a more refined estimation based on mixing times \cite{levin_markov_2017} is possible, allowing one to control closeness to stationarity to an arbitrary level of precision.

\subsubsection{Number of Elements in the Queue} \label{subsubsec:number_events}
The number of elements that join the queue in a time interval $[t_1, t_2]$ is, by design, Poisson distributed with average $\lambda (t_2-t_1)$. The number of elements in the queue in the same interval is slightly different. In addition to new elements joining in that interval, we need to count how many started before and ended after $t_1$. This amounts approximately to the sum of two independent random variables, one Poisson with average $\lambda (t_2 - t_1)$ and the other Poisson with average $m(t_1)$, which by superposition is again Poisson with average $\lambda (t_2 - t_1) + m(t_1)$. In the stationary regime, this becomes a Poisson random variable with average $\lambda (t_2 - t_1) + \lambda/\mu$.
We use these facts later in \cref{subsec:generate_snapshots}.

\section{The Ascona Model and EDLDE Structures} \label{sec:Ascona}
This section introduces a link stream model and a practical sampling procedure.
In the absence of a previously established name, we refer to it as the \emph{Ascona model} for clarity and convenience.
To avoid ambiguity, we explicitly distinguish the \emph{model} (a probability distribution on link streams)
from the \emph{sampler} (an algorithm that draws approximately from that distribution).

\subsection{The Ascona Model}
Let $V=\{v_1,\dots,v_N\}$ be a vertex set and $T\subset\mathbb{R}$ a time domain.
Following \cref{subsubsec:linkstreams}, a (simple) link stream is a triplet
\[
L=(T,V,E), \qquad 
E\subseteq \bigl\{([\alpha,\omega),\{u,v\}) : \alpha \le \omega,\ u\neq v \bigr\},
\]
so that $([\alpha,\omega),\{u,v\})\in E$ means that the link $\{u,v\}$ is present for all $t\in[\alpha,\omega)$.
We denote by
\begin{equation}
    M_L(t) \;\coloneqq\; \bigl|\{\{u,v\} : ([\alpha,\omega),\{u,v\})\in E,\ t\in[\alpha,\omega)\}\bigr|
\end{equation}
the number of active links at time $t$.

A convenient way to specify and sample $E$ is via a marked point process of the link start times.
Let $\{T_k\}_{k\ge 1}$ be random start times on $T$, and let marks be $(D_k,\{U_k,V_k\})$ with $D_k>0$ and $U_k\neq V_k$.
Each $T_k$ generates one element
\[
([T_k,T_k+D_k),\{U_k,V_k\})\in E.
\]
This representation is equivalent to the interval-based definition above, but is more convenient for the Ascona model construction.

The model generates a random link stream by combining:
\begin{enumerate}
    \item a \emph{temporal mechanism} producing a (multi)set of link intervals, and
    \item a \emph{connectivity mechanism} assigning an unordered pair of nodes to each interval.
\end{enumerate}

The Ascona model assumes that \emph{when} and \emph{for how long} links occur is decoupled from \emph{which} node pairs interact.
Formally, the endpoint marks are independent of the temporal marks:
\begin{equation} \label{eq:Ascona_assumption}
    \{\{U_k,V_k\}\}_{k\ge 1} \ \perp\!\!\!\perp\ \{(T_k,D_k)\}_{k\ge 1}.
\end{equation}

Equivalently, conditional on the realized collection of intervals
$\mathcal{I}=\{[T_k,T_k+D_k)\}_{k\ge 1}$, the unordered pairs $\{U_k,V_k\}_{k\ge 1}$ are i.i.d.\ draws from a fixed distribution on $\binom{V}{2}$
that does not depend on $\mathcal{I}$.
We now describe both mechanisms in more detail.

\subsubsection{Temporal Mechanism}
\label{subsec:ascona_temporal}
In the Ascona model, the temporal structure is defined by a queueing process $A/S/c/K/\bar{K}/D$ on a base time interval $[0,t_e]$,
followed by its tail, obtained by stopping admissions after $t_e$ while letting in-service customers complete.
A realized queue trajectory produces a collection of active intervals
$\mathcal{I}=\{[T_k,E_k)\}$ (``customers'') and therefore an occupancy process $M(t)$ (number of simultaneously active customers).
Here $[T_k,E_k)$ corresponds to the generic interval notation $[\alpha,\omega)$ used in \cref{subsubsec:linkstreams}.

In practice, one often selects only a portion of the queue trajectory (e.g. head, tail, or an approximately stationary segment),
as described later in \cref{subsubsec:archetypes}.
Additionally, a block can be placed anywhere in the global time domain by applying a time shift $\tau_a$ (defined in eq. \eqref{eq:shift}).
In this work, we treat trimming and placement as user-chosen design decisions; one could alternatively randomize them.

\subsubsection{Connectivity Mechanism}
\label{subsubsec:ascona_connectivity}
By the Ascona decoupling assumption (eq. \eqref{eq:Ascona_assumption}), after sampling link intervals, endpoints are assigned stochastically and independently across events,
according to an endpoint law that is independent of the temporal realization.
We consider two equivalent parametrizations of the endpoint law on unordered pairs $\binom{V}{2}$.

\paragraph{Connectivity probability matrix (one-step)}
Let $\pi$ be a probability distribution on unordered pairs $\binom{V}{2}$.
We encode $\pi$ by a symmetric matrix $P\in[0,1]^{N\times N}$ with $P_{ii}=0$ and
\begin{equation}\label{eq:P_normalization}
    P_{ij}=P_{ji},\qquad \sum_{1\le i<j\le N} P_{ij}=1.
\end{equation}
In this representation, for $i<j$ we have $\pi(\{v_i,v_j\})=P_{ij}$.
We call $P$ the \emph{connectivity probability matrix}.

\paragraph{Connectivity conditional probability matrix (two-step)}
Alternatively, sample a first endpoint $U\sim p$ for a distribution $p$ on $V$,
then sample a second endpoint $V\sim Q_{U,\cdot}$ where $Q$ is row-stochastic with $Q_{ii}=0$.
This factorization is convenient to separate heterogeneous activity ($p$) \cite{perra_activity_2012} from mixing patterns ($Q$).

The pair $(p,Q)$ induces a distribution on unordered pairs by symmetrization:
\begin{equation}\label{eq:pi_from_pQ}
    \pi(\{v_i,v_j\})
    \;=\;
    \frac{1}{2}\Big(p_i\,Q_{ij}+p_j\,Q_{ji}\Big).
\end{equation}
The one-step matrix $P$ corresponding to $(p,Q)$ is given by
$P_{ii}=0$ and, for $i<j$, $P_{ij}=\pi(\{v_i,v_j\})$ (hence $P_{ij}=P_{ji}$).

A convenient choice for $p$ is the uniform distribution, which has maximal entropy and therefore can be interpreted as uninformative from a Bayesian perspective.

The practical advantage of the Ascona framework is that complex, nonstationary activity profiles can be built
by composing simpler sub-stream units called queue blocks.
This makes model design modular: temporal structure is engineered at the block level and then combined together.

\subsubsection{Ascona Sampler: Queue Blocks}
\label{subsubsec:ascona_blocks}
A \emph{queue block} $B$ is a link stream sampled on a base interval $[0,t_e]$ in two stages:
\begin{enumerate}
    \item \textbf{Temporal stage:} sample a collection of intervals $\mathcal{I}_B=\{[T_k,E_k)\}_{k=1}^{n_B}$ from the chosen queue model,
    optionally trimmed (head/tail/stationary selection).
    \item \textbf{Connectivity stage:} for each $[T_k,E_k)\in\mathcal{I}_B$, sample an unordered pair $\{U_k,V_k\}$ i.i.d.\ from the chosen endpoint law
    (either $\pi$ or $(p,Q)$), independently of $\mathcal{I}_B$.
\end{enumerate}

For a shift $a\in\mathbb{R}$ define the time shift operator
\begin{equation} \label{eq:shift}
    \tau_a(L) \coloneqq (T+a, V, E_a), \qquad
E_a \coloneqq \{([\alpha+a,\omega+a),\{u,v\}) : ([\alpha,\omega),\{u,v\})\in E\}.
\end{equation}

Given two streams $L_1=(T_1,V,E_1)$ and $L_2=(T_2,V,E_2)$ on the same node set, define their \emph{superposition}
\begin{equation}
    L_1\oplus L_2 \;\coloneqq\; (T_1\cup T_2,\ V,\ E_1\cup E_2),
\end{equation}
interpreting $E_1\cup E_2$ either as a set (simple stream) or as a multiset (allowing duplicates), depending on the application.
\emph{Concatenation} is the special case where the time supports are disjoint.

If blocks $B_1,\dots,B_R$ are placed at times $a_1,\dots,a_R$, the global stream is
\begin{equation}
    L \;=\; \bigoplus_{r=1}^R \tau_{a_r}(B_r).
\end{equation}

\subsubsection{Conflicts}
\label{subsec:ascona_conflicts}
A conflict arises whenever we attempt to add a link that overlaps in time with an already present link on the same unordered pair.
Concretely, a proposed link $\ell_2=([s_2,e_2),\{u,v\})$ conflicts with an existing link $\ell_1=([s_1,e_1),\{u,v\})$
if $s_2<e_1$ and $s_1<e_2$.

If multi-links are allowed (i.e. $E$ is treated as a multiset), the sampler draws exactly from the intended construction.
If we enforce simplicity (at most one active link per pair at any time), conflict handling can be interpreted as a \emph{projection}
from an ideal multiset-valued sample to the space of simple link streams, and the sampler becomes approximate.

When looking at a single queue block, there are two main causes for conflicts:
\paragraph{Congestion / saturation}
A simple link stream on $N$ nodes admits at most $\binom{N}{2}$ simultaneous active links.
The temporal mechanism may still be specified by a queue with an abstract capacity $K>\binom{N}{2}$, but then enforcing simplicity requires a projection step.
We can estimate if a block is prone to this problem by looking at the typical occupancy level of the underlying queue: if it is well below $\binom{N}{2}$, such projections are rarely activated. 

\paragraph{Endpoint concentration}
Another cause for conflicts is due to connectivity constraints. For instance, if a node $v_i$ is already connected to $v_j$ and the endpoint law forces new links out of $v_i$ to connect almost exclusively to $v_j$, then a new proposed interval is likely to target $\{v_i,v_j\}$ again.
More generally, if the endpoint law heavily favors a small subset of pairs, resampling an already active pair becomes likely even far from saturation.
We can gauge how problematic a certain connectivity probability matrix is by computing its Shannon entropy. Similarly, to gauge a connectivity conditional probability matrix, we can compute its conditional entropy conditioned on the sampling vector; analogously to \cite{koovely_evolution_2025}, where the probabilistic matrix is a transition operator between nodes rather than a connectivity one.

Possible projection rules include:
\begin{itemize}
    \item \textbf{Merge:} replace $\ell_1$ and $\ell_2$ by $([\min(s_1,s_2),\max(e_1,e_2)),\{u,v\})$,
    \item \textbf{Discard:} ignore $\ell_2$ and keep $\ell_1$,
    \item \textbf{Resample endpoints:} keep the interval $[s_2,e_2)$ but resample the unordered pair $\{u,v\}$.
\end{itemize}
Each rule introduces small, local deviations from the target temporal/connectivity laws. One might choose a rule to mitigate discrepancies from one or the other aspect, or based on practical considerations: for instance, if a queue is saturated, resampling nodes won't help.
Importantly, the presence of an error does not increase the probability of another error occurring, so deviations do not propagate, and actually, there is a self-adjustment mechanism: once the overlapping interval between conflicting links ends, the discrepancy disappears.

Superposition of blocks can cause conflicts as well and can be resolved with the same projection rules.
We summarize the sampling procedure in \cref{alg:ascona_sampler}.

\begin{algorithm}[ht]
\caption{Ascona sampler (block-based)}
\label{alg:ascona_sampler}
\begin{algorithmic}[1]
\REQUIRE Vertex set $V$; blocks $(\mathsf{Queue}_r,\ t_e^{(r)},\ a_r,\ \text{trim}_r,\ \text{endpoint-law}_r)_{r=1}^R$; conflict rule $\mathsf{Resolve}$
\ENSURE Link stream $L=(T,V,E)$
\STATE $E\leftarrow \emptyset$
\FOR{$r=1$ \TO $R$}
    \STATE Sample intervals $\mathcal{I}_r=\{[T_k,E_k)\}_{k=1}^{n_r}$ from $\mathsf{Queue}_r$ on $[0,t_e^{(r)}]$, apply $\text{trim}_r$
    \FOR{each $[T_k,E_k)\in \mathcal{I}_r$}
        \STATE Sample unordered pair $\{U_k,V_k\}$ from $\text{endpoint-law}_r$ \COMMENT{independent of $\mathcal{I}_r$}
        \STATE Propose link $\ell=([T_k+a_r,E_k+a_r),\{U_k,V_k\})$
        \STATE $E\leftarrow \mathsf{Resolve}(E,\ell)$ \COMMENT{projection step}
    \ENDFOR
\ENDFOR
\STATE $T\leftarrow \bigcup_{r=1}^R [a_r,a_r+t_e^{(r)}]$
\RETURN $L=(T,V,E)$
\end{algorithmic}
\end{algorithm}

\begin{figure}[htbp]
  \centering
    \includegraphics[width=0.95\linewidth]{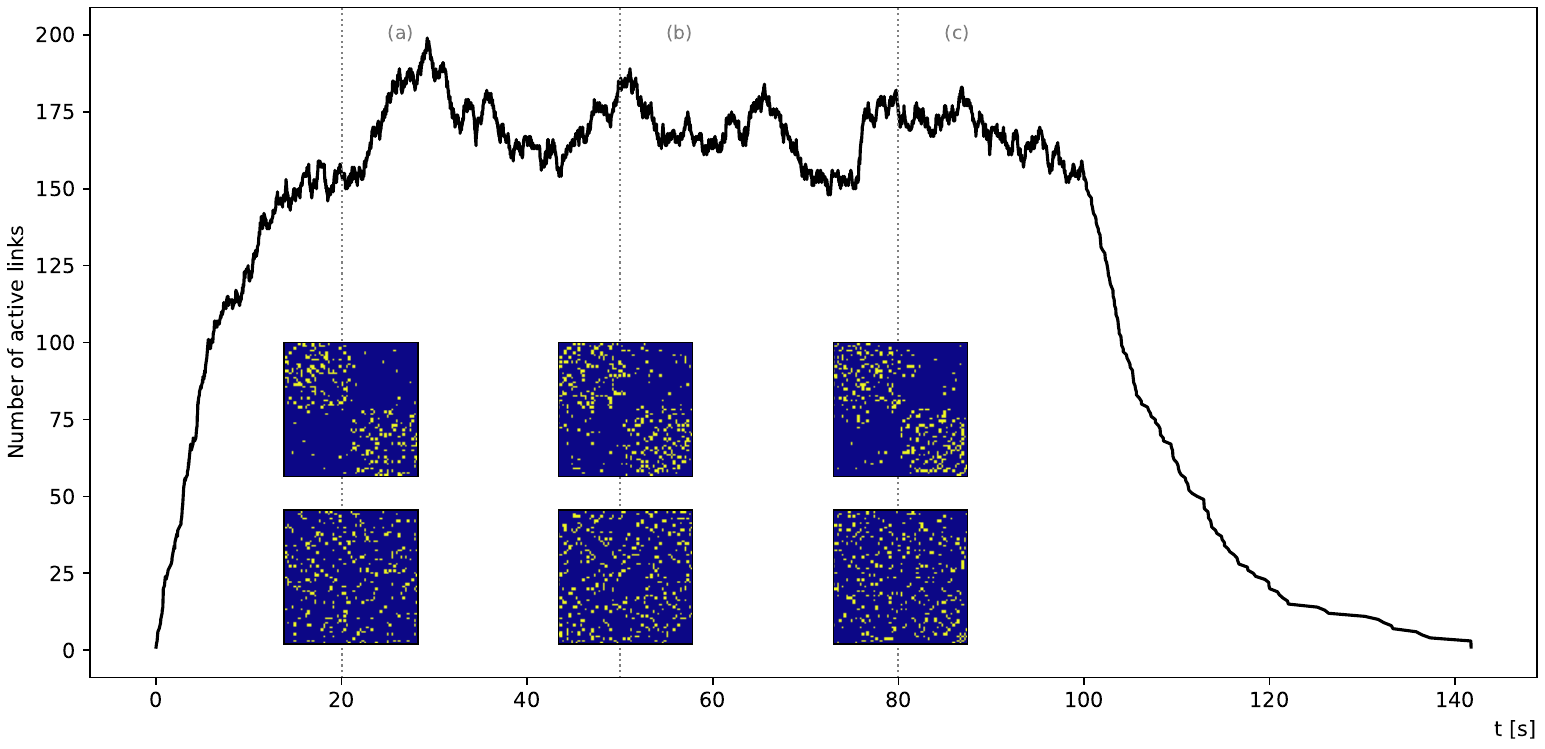}
  \caption{Example of queue blocks $L'$ and $L''$ sharing the same links' repartition in time (sampled from a $M/M/\infty$ process), but different connectivity structures. For each of the two sampled link streams, we display the adjacency matrix of the instantaneous graphs $G_t'$ and $G_t''$ at three different times $(t \in \{20,50,80\})$. On top, we show the ones of $L'$, which has a constant two-block connectivity structure. On the bottom side, we show the ones of $L''$, which does not have any particular structure: all pairs of nodes have the same probability to be connected, hence this could be considered a link stream version of an ER graph.}
  \label{fig:fig_EDLDE_comparison}
\end{figure}

\begin{figure}[htbp]
  \centering
    \includegraphics[width=0.95\linewidth]{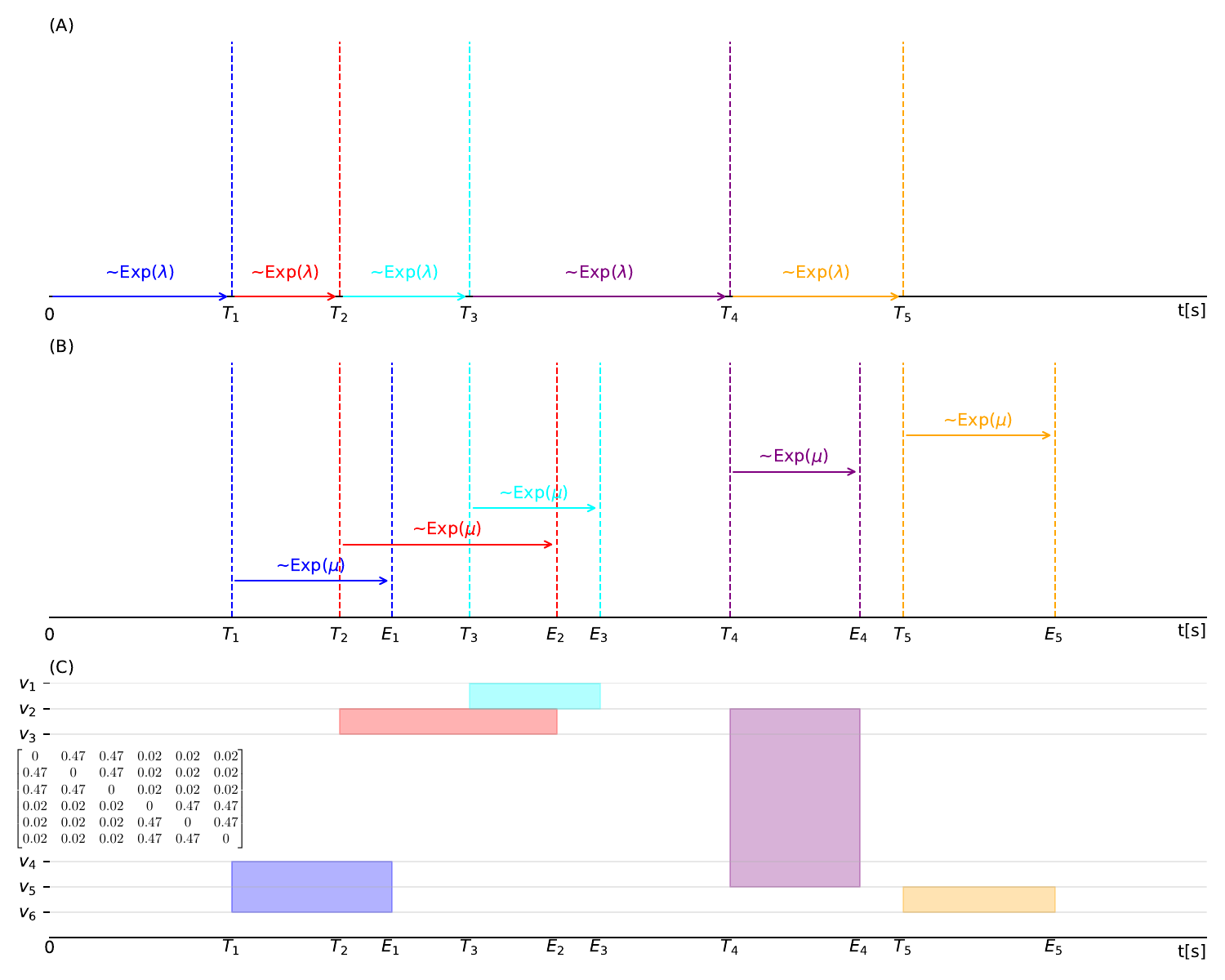}
  \caption{Representation of the generation of an EDLDE structure with the Ascona model. (A) The first step consists of drawing i.i.d.\ exponentially distributed inter-arrival times with rate $\lambda$, or equivalently sampling start times from a Poisson process with rate $\lambda$, on a desired time interval. Each sampled time corresponds to a starting time $T_i$ of a link; in this example, there are five of them: $T_1, \dots, T_5$. (B) The second step consists of sampling an i.i.d. exponential duration with rate $\mu$ for each start time. In this example, we obtain five link intervals $[T_1,E_1), \dots, [T_5,E_5)$. (C) Finally, each interval is assigned an unordered pair of vertices. For each link, the first endpoint is drawn uniformly at random from the node set and the partner is drawn from the connectivity conditional probability matrix displayed on the left. In this case, the matrix encodes a two-block structure with high within-block probability and low between-block probability. We display node names on a vertical axis on the left of the figure. Each link is represented by a colored rectangle showing the involved nodes and the time interval when it occurs.}
  \label{fig:EDLDE_Descritiption}
\end{figure}

We show a concrete example of two dynamic networks sampled with the Ascona sampler in \cref{fig:fig_EDLDE_comparison}.
The two samples share the same temporal structure but differ in their endpoint laws.
The shared temporal structure is sampled based on an $M/M/\infty$ queue.
We now discuss in more detail this fundamental parametrization, whose construction is illustrated in \cref{fig:EDLDE_Descritiption}.

\subsection{EDLDE Structures} \label{subsec:EDLDE}
An EDLDE structure is a link stream generated based on the following inputs: a starting time $t_s$, an ending time $t_e$,
an activity rate $\lambda$, and an interaction rate $\mu$.
We assume that link start times form a homogeneous Poisson process on $[t_s,t_e]$ with rate $\lambda$.
Equivalently, the number of starts in $[t_s,t_e]$ is $\mathrm{Poisson}(\lambda (t_e-t_s))$.
Additionally, link durations are i.i.d. exponential with rate $\mu$.
We therefore have \textbf{E}xponential-\textbf{D}uration \textbf{L}inks \textbf{D}istanced \textbf{E}xponentially, hence the acronym EDLDE.

We can conveniently sample such structures with the Ascona method.
The underlying temporal mechanism corresponds to an $M/M/\infty$ queue, optionally capped, as in  $M/M/\binom{N}{2}/\binom{N}{2}/\infty/\mathrm{FCFS}$ \cite[Ch.\ 2.8]{gross_fundamentals_2008},
a truncated counterpart of $M/M/\infty$ \cite[Ch.\ 6.7]{robert_stochastic_2013}.
In the regime $\binom{N}{2} \gg \frac{\lambda}{\mu},$ the occupancy is far from saturation, so distortions due to the finite-pair constraint are boundary effects and the temporal mechanism is well approximated by the $M/M/\infty$ queue.

From an information-theoretical or Bayesian perspective, an EDLDE structure can be seen as a reference null model that does not assume any knowledge of any correlation between arrival and service times:
exponential distributions are uninformative priors because they have maximum differential entropy among all continuous distributions supported on $[0, \infty)$ with the same mean \cite{dowson_maximum-entropy_1973}.

However, one might be interested in sampling links based on other distributions than exponential ones.
In that case, we obtain queues alternative to the $M/M/\infty$ one.
Their implementation leads to small modifications of the Ascona sampling method and produces structures different from the EDLDE ones.
Theoretical properties of these derived models depend on the corresponding properties of the alternative queueing processes; however, some properties, like the stationary mean occupancy, are often independent of the purely Markovian assumptions \cite{bunday_gmr_1980, gross_fundamentals_2008}.
See \cite{gross_fundamentals_2008} for a discussion of the consequences of dropping the Markovian assumption for the arrival process, in particular allowing restrictions on the support, inhomogeneity of the rate parameter, and batch arrivals.

\subsubsection{Degenerate Cases}
We can generate link streams with 0-duration links with the Ascona model by setting the service times to be Dirac $\delta$ distributed. In the EDLDE structure case, this reduces to a system with links spawning as a Poisson process. \\
Instances of discrete-time link stream can be sampled by picking arrival times, and edge durations according to $\mathbb{N}$-supported distributions. 

\subsubsection{Smooth Stochastic Blocks} \label{subsubsec:SSBM}
In this part, we describe how to use the Ascona model to sample link streams with evolving community structures. The idea is to concatenate blocks, or equivalently, to design a single queue block whose connectivity (conditional) probability matrix evolves as a function of time and follows the desired clusters' evolution. In this article, we show EDLDE samples exclusively, but the desired effect is achievable almost independently of the underlying queue. 

In the example displayed in \cref{fig:fig_EDLDE_comparison}, we used constant two-block connectivity conditional probability matrices, leading to a constant community structure. In the example displayed in \cref{fig:fig_EDLDE_events}, we first have a four-block structure followed by a two-block structure. The change represents a discontinuity in the progression of the underlying connectivity conditional probability matrix; however, preexisting links take a while to disappear after the change, while links associated with the new structure appear gradually, leading to smooth transitions, hence the name ``smooth stochastic blocks".

In \cref{subsec:generate_snapshots}, we explain how to aggregate smooth stochastic blocks in order to get snapshots, and after that, we compare them to Dynamic SBMs, the main generalization of SBMs to temporal networks. Now, we showcase the expressivity of the newly introduced smooth SBMs.

\subsubsection{Generating Archetypes} \label{subsubsec:archetypes}
A fundamental aspect of temporal networks is the understanding of their clusters and how they evolve in time. Similarly to the methods used in \cite{asgari_mosaic_2023} and \cite{cazabet_evaluating_2021}, the procedure to generate link stream scenarios follows two steps: first, the experimenter needs to describe the scenario in terms of a queue block's footprint, and secondly, generate the structure with the Ascona model to obtain a temporal network that qualitatively reflects the desired scenario.

In this part, we display the expressivity of the Ascona model by describing how to generate the typical ``events" characterizing temporal networks' evolution listed in \cite{cazabet_challenges_2023}. For elementary versions of such scenarios, a single EDLDE queue block is sufficient. In order to design some of these events, one can focus exclusively on the temporal aspect of the model.

\paragraph{Birth/Death}
In a birth event, a new community appears. Conversely, in a death event, an existing community vanishes.
One can model the birth of a community by either taking the head or the time-reversed tail of an EDLDE queue block. In the first case, the growth is not necessarily monotonic and has roughly a concave shape. In the second case, the growth is monotonic and has approximately a convex shape. Conversely, one can model the death of a community by taking the time-reverse of a birth event.

\paragraph{Continuity}
One can model a stable community by taking a queue block in its stationary regime.

\begin{figure}[htbp]
  \centering
    \includegraphics[width=0.95\linewidth]{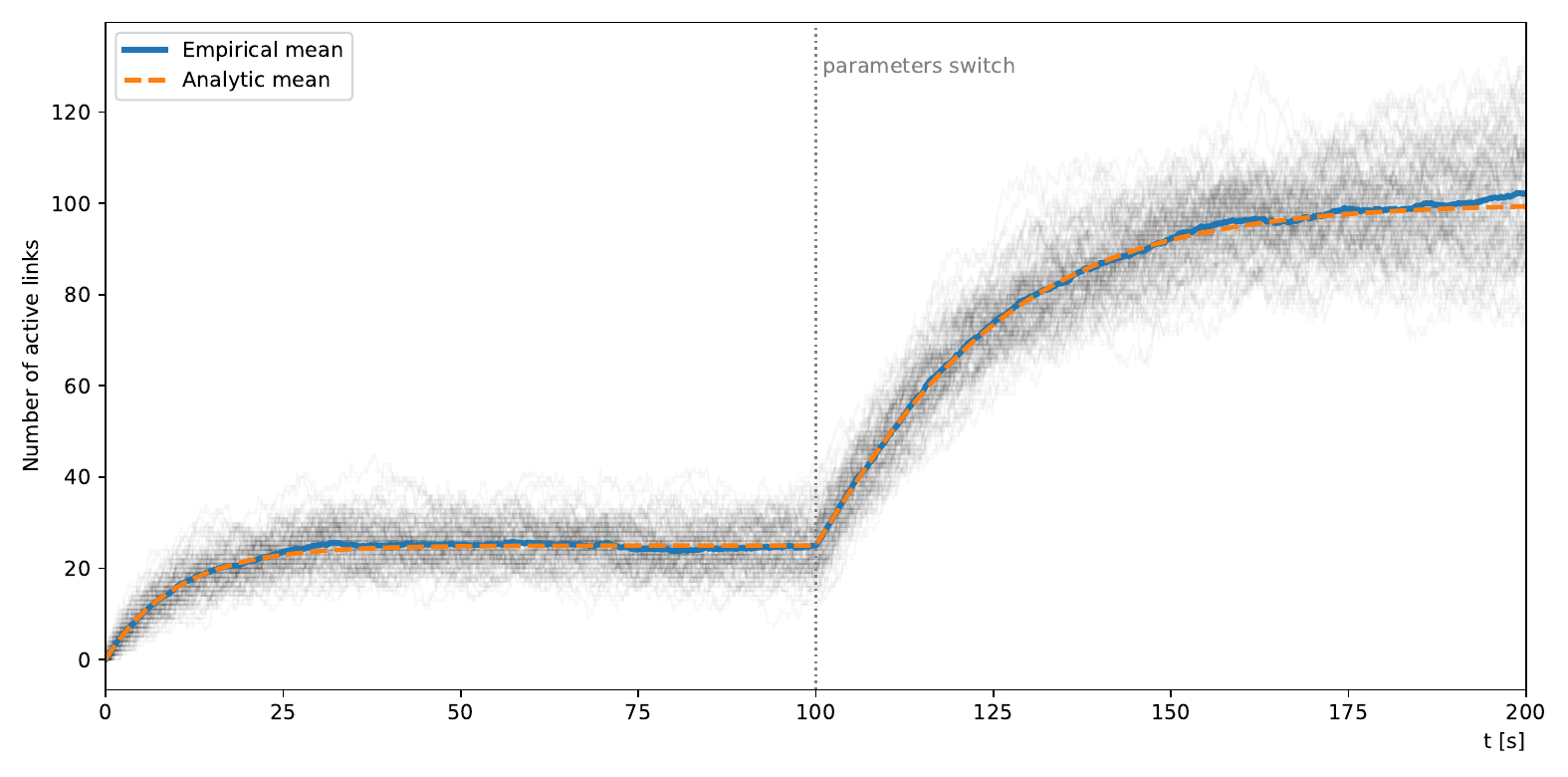}
  \caption{Trajectories of 100 samples of a $M/M/ \infty$ queue with changing parameters at time $t_{sw} = 100 \,s$. Before $t_{sw}$, the rate of arrivals is $\lambda_1 = 2.5$ and the service time rate $\mu_1 = 0.1$; after $t_{sw}$, they are equal to $\lambda_2 = 5$, and $\mu_2=20$. We compare the sample mean with the analytical mean from eq. \eqref{eq:switch_m} until the time $t_e=200$ when we stop new elements from joining the queue.}
  \label{fig:trajectories_chain}
\end{figure}

\paragraph{Change of intensity}
One event, not listed among the ones in \cite{cazabet_challenges_2023}, is the one where a stable community experiences an increase/decrease in the number of interactions. We can model this situation with an EDLDE queue block by imposing that the parameters $\lambda$ and $\mu$ of the underlying $M/M/ \infty$ process change at a switching time point $t_{sw}$. The underlying process remains Poissonian, and its average is
\begin{equation} \label{eq:switch_m}
    \resizebox{.90\hsize}{!}{$m(t)=
    \begin{cases}
    0, & t<t_{s},\\[6pt]
    \frac{\lambda_1}{\mu_1}\!\left(1-e^{-\mu_1 (t-t_{s})}\right),
    & t_{s}\le t\le t_{sw},\\[8pt]
    \frac{\lambda_1}{\mu_1}\!\left(e^{-\mu_1 (t-t_{sw})}-e^{-\mu_1 (t-t_{s})}\right)
    \;+\;
    \frac{\lambda_2}{\mu_2}\!\left(1-e^{-\mu_2 (t-t_{sw})}\right),
    & t_{sw}\le t\le t_e,\\[8pt]
    \frac{\lambda_1}{\mu_1}\!\left(e^{-\mu_1 (t-t_{sw})}-e^{-\mu_1 (t-t_{s})}\right)
    \;+\;
    \frac{\lambda_2}{\mu_2}\!\left(e^{-\mu_2 (t-t_e)}-e^{-\mu_2 (t-t_{sw})}\right),
    & t>t_e
    \end{cases}$}
\end{equation}
We display in \cref{fig:trajectories_chain} the quality of this formula with simulations of $M/M/ \infty$ queue experiencing an increase in activity due to changing parameters.

More complex temporal scenarios require more complex concatenations.
\paragraph{Resurgence}
 Resurgence can be thought of as a combined event where a queue block vanishes for a period, then reappears without particular perturbations. \\

Other events, the ones that have more to do with how communities interact, can be designed by intervention on the connectivity aspects of a queue block.

\paragraph{Merge/Split}
A merge event, as the name says, indicates that multiple existing clusters of nodes merge into one. Conversely, in a split, an existing cluster divides into multiple smaller clusters. One can easily engineer such events with a queue block that follows a certain connectivity matrix before the merge/split event, and a new one after.

\paragraph{Growth/Contraction}
The growth and contraction events are similar to the previous case. In a growth event, we observe an existing community acquires new nodes; in a contraction, a community loses some nodes, but does not disappear completely. In these cases, instead of two interacting communities, one community interacts with a set of unstructured, lowly connected nodes.
\\

One can notice that this procedure leads to controllably smooth transitions, balancing stability in time of the links with consistency with the connectivity structures, a known problematic aspect of sampling temporal networks \cite[Section 3.2.1]{cazabet_evaluating_2021}.

\begin{figure}[htbp]
  \centering
    \includegraphics[width=0.95\linewidth]{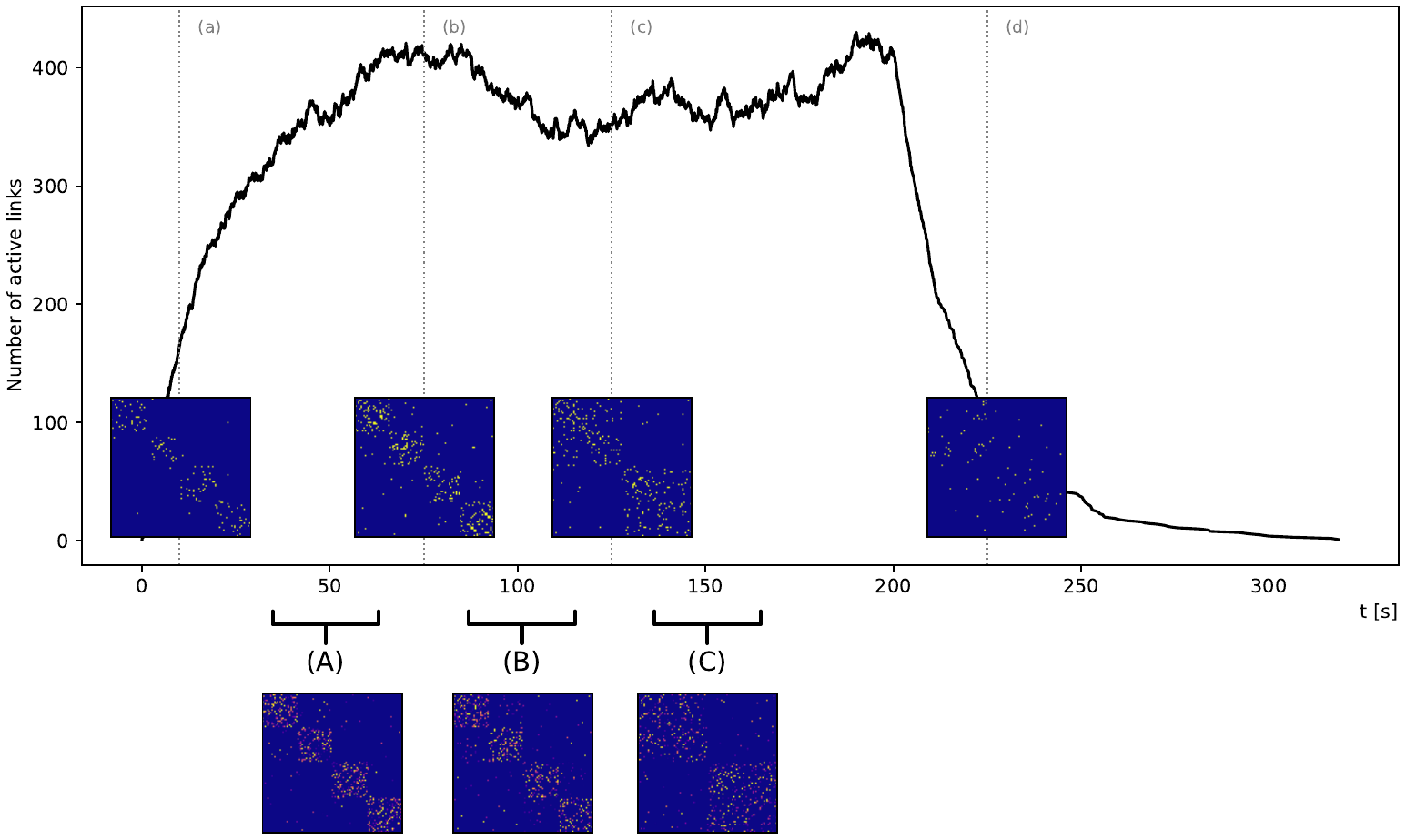}
  \caption{Ascona model sample displaying various events. (a) Birth event of four communities. (b) In the stationary regime, there is a stable configuration with four fully developed communities. (c) At time 100, there is a merge event leading to the formation of two bigger communities. (d) After time 200, the queue decays, and the two communities vanish. (A) Weighted footprint of the network of the interval $[40 \,s,60 \,s]$ displaying clearly the four-block structure before the merge. (B) Weighted footprint of the interval $[90 s,110s]$ displaying the transition between the four and the two-block structure happening at time $100s$. (C) Weighted footprint of the interval $[140 s,160s]$ displaying the two-block structure after the merge. The colour of each footprint entry depends on the weight from eq. \eqref{eq:footbprint_w}. Intuitively, warmer colours indicate higher weights.}
  \label{fig:fig_EDLDE_events}
\end{figure}

We display in \cref{fig:fig_EDLDE_events} the generation of many of these events with a single EDLDE block, encoding the evolution of a link stream from a four-block to a two-block structure. We show as well that by slicing the time domain and aggregating the information over the slices, we get a snapshot equivalent of this link stream. We discuss this procedure in more detail in the next part.  

\subsection{Generating Snapshot Networks} \label{subsec:generate_snapshots}
As mentioned previously, the majority of datasets and methods in temporal networks theory have been designed primarily (and often exclusively) with snapshot networks in mind. We therefore describe a procedure for generating snapshot networks from an EDLDE sample in a controlled way. The strategy is to discard the head of the queue based on the estimated crossover times (eq. \eqref{eq:tstar_exact} or its approximation eq. \eqref{eq:tstar_approx}) presented in \cref{subsec:head_queue}, and then slice the time domain and aggregate the links in unweighted footprints (see eq. \eqref{eq:footprint_u}), which leads to ``uniform" snapshots.

We suggest two uniformity criteria for generating homogeneous snapshots. The first criterion is to try to produce snapshots from time intervals with the same length $\Delta t$, so that they contain the same average number of spawning links. Let this number be $K$; then, from \cref{subsubsec:number_events}, we know that in the stationary regime, 
\begin{equation}
    \lambda \Delta t \overset{!}{=} K,
\end{equation} which leads to 
\begin{equation} \label{eq:criterion_links_spawning}
    \Delta t = \frac{K}{\lambda}.
\end{equation}
Another uniformity criterion is to impose $K$ to be the total number of active links, on average, in each snapshot.
In this case, always from \cref{subsubsec:number_events}, we know that, in the stationary regime, we have
\begin{equation}
    \lambda \Delta t  + \frac{\lambda}{\mu}  \overset{!}{=} K,
\end{equation} which leads to 
\begin{equation} \label{eq:criterion_links_total}
    \Delta t = \frac{K}{\lambda} - \frac{1}{\mu}.
\end{equation}

In both cases, one could increase the overlap between subsequent slices to achieve a higher level of smoothness.
Notice that if one wants to exert control on the number of distinct links in every slice, then a correction factor for eq. \eqref{eq:criterion_links_spawning} and eq. \eqref{eq:criterion_links_total} is required to take into account the probability of links appearing multiple times in a certain $\Delta t$.

\subsubsection{Comparison with Dynamic SBM}
If we apply the above-presented procedure to smooth SBM link streams, we obtain snapshots with block structures. This is reminiscent of the  Dynamic SBM model, which has emerged as a popular extension of its static network counterpart and is used, for instance, for testing change-point detection methods (e.g., in \cite{bhattacharjee_change_2020, sulem_graph_2024}). In dynamic SBMs, each snapshot is sampled according to a static SBM distribution whose parametrization can change in time. In terms of expected number of edges and its variance, smooth SBMs and dynamic SBMs are similar because of the similarity between Poisson and Binomial distributions. The two methods differ in terms of smoothness: dynamic SBM snapshots have no mechanism to enforce smoothness, while the smoothness of the underlying smooth SBM link stream is inherited by the derived SBM snapshots.

\section{Conclusions} \label{sec:Conclusions}
This article focused on sampling link streams with non-zero duration links, but throughout the work, we highlighted steps where minor additions or modifications would lead to different possibilities, such as multi/self/weighted/signed links. Additionally, we discussed how to obtain other types of temporal graphs: discrete-time link streams, 0-duration link ones, and snapshot networks. Graph streams are more complex than link streams, and one requires an extra ingredient to sample them compared to link streams: a sampling mechanism for the temporal presence of each node.

In \cref{subsubsec:SSBM}, we introduced smooth SBM, a specific variation of the Ascona model that generates link streams with community structures, and compared it with the dynamic SBM model. Similarly, one could design alternative (conditional) probability connectivity matrices to generate structures with different properties, e.g., nestedness, disassortativity, and structurelessness.
In \cref{subsubsec:archetypes}, we have seen how to generate archetypical temporal structures with the smooth SBMs. As discussed in \cite{failla_describing_2024},  hybridization of these simple archetypes leads to more complex structures. It is realistic to imagine generating hybrids by combining several queue blocks, and this is a possible future direction of this work, raising the question of how expressive the Ascona model really is.

Aside from these positive aspects of expressivity and flexibility, a big advantage of this article's approach is that it translates temporal graphs' problems to problems of time series and stochastic processes that are well studied and understood, with a rich body of theoretical and practical results ready to be exploited. For instance, \cite{walrand_introduction_1988} discusses various aspects of statistical analysis on queues. $M/M/\infty$ models, the foundation of the EDLDE structures, are particularly nicely behaved from a statistical perspective: for instance, the MLE estimation is equivalent to the method of moments when fitting the exponential distributions of the model.
Other queue models would require the use of simulations, but some are remarkably amenable to analyze (e.g., see \cite{eick_physics_1993}).

Future work should address aspects of inferential statistics: given a reference link stream, find the Ascona model parametrization that fits the data in the best way possible. Recent works on maximum entropy approaches to temporal networks \cite{barucca_maximum_2025, clemente_temporal_2024} could pave the way in this direction and go beyond the purely Markovian framework.

\section*{Material}
A GitHub project containing the code used for the generation of all plots is available\footnote{\url{https://github.com/samuelkoovely/EDLDE_project}}.

\appendix
\section{Head of the Queue: Computations} \label{sec:appendix_head} Here we derive the crossover time $t^*$ as defined in eq. \eqref{eq:criterion}.

Substituting eq. \eqref{eq:delta_def} and eq. \eqref{eq:sigma_def} into
eq. \eqref{eq:criterion} gives
\begin{equation}
\frac{\lambda}{\mu}  e^{-\mu t^\star} 
=
\sqrt{\frac{\lambda}{\mu} \left(1 - e^{-\mu t^\star}\right)}.
\label{eq:main_equation}
\end{equation}

Let
\begin{equation}
\rho \coloneqq \frac{\lambda}{\mu},
\qquad
x \coloneqq e^{-\mu t^\star},
\quad \text{so that} \quad
t^\star = -\frac{\ln (x)}{\mu}.
\end{equation}
Then eq. \eqref{eq:main_equation} becomes
\begin{equation}
\rho x = \sqrt{\rho \left(1-x\right)}.
\end{equation}
Squaring both sides yields
\begin{equation}
\rho^2 x^2 = \rho(1-x)
\quad\Longrightarrow\quad
\rho x^2 + x - 1 = 0.
\label{eq:quad}
\end{equation}
Equation eq. \eqref{eq:quad} is quadratic in $x$ with solutions
\begin{equation}
x =
\frac{-1 \pm \sqrt{1 + 4\rho}}{2\rho}.
\end{equation}
Because $x = e^{- \mu t^\star}$ must be positive, we take the `$+$' root:
\begin{equation}\label{app:x_exact}
x
=
\frac{-1 + \sqrt{1 + 4\rho}}{2\rho}.
\end{equation}
Finally,
\begin{equation} \label{app:tstar_exact}
t^\star
=
-\frac{1}{\mu}
\ln\!\left(
\frac{-1 + \sqrt{1 + 4 \frac{\lambda}{\mu}}}{2 \frac{\lambda}{\mu}}
\right).
\end{equation}
Equation eq. \eqref{app:tstar_exact} is an exact expression for the crossover time $t^\star$ at which the deterministic distance from stationarity
falls to the same order as the instantaneous stochastic fluctuations.

We can approximate this value when the steady-state mean load $\rho$ is large enough ($\rho \gg 1$). In that case we can simplify eq. \eqref{app:x_exact} using
$-1 + \sqrt{1+4\rho} \approx 2\sqrt{\rho}$:
\begin{equation}
x
\approx
\frac{2\sqrt{\rho}}{2\rho}
=
\frac{1}{\sqrt{\rho}},
\qquad\text{so}\qquad
t^\star \approx
- \frac{1}{\mu} \ln \left(\frac{1}{\sqrt{\rho}}\right)
=
\frac{1}{2\mu} \ln(\rho).
\end{equation}
Thus, for moderately large $\frac{\lambda}{\mu}$, an accurate rule of thumb is
\begin{equation} \label{app:tstar_approx}
    t^\star \approx \frac{1}{2 \mu} \ln(\frac{\lambda}{\mu}).
\end{equation}

\section*{Acknowledgments}
Y. Asgari and G. Vaccario suggested articles relevant to the subject, while A. Bovet and Z. Fang's comments helped me to improve this manuscript. C. Casati helped in finding the catchy palindromic acronym EDLDE. My gratitude goes to all of them.

\bibliographystyle{amsplain}  
\bibliography{refs}  

@article{bovet_flow_2022,
	title = {Flow stability for dynamic community detection},
	volume = {8},
	url = {https://www.science.org/doi/abs/10.1126/sciadv.abj3063},
	doi = {10.1126/sciadv.abj3063},
	number = {19},
	journal = {Science Advances},
	author = {Bovet, Alexandre and Delvenne, Jean-Charles and Lambiotte, Renaud},
	year = {2022},
	note = {\_eprint: https://www.science.org/doi/pdf/10.1126/sciadv.abj3063},
	pages = {eabj3063},
}

@article{abbe_community_2018,
	title = {Community {Detection} and {Stochastic} {Block} {Models}: {Recent} {Developments}},
	volume = {18},
	url = {http://jmlr.org/papers/v18/16-480.html},
	number = {177},
	journal = {Journal of Machine Learning Research},
	author = {Abbe, Emmanuel},
	year = {2018},
	pages = {1--86},
}

@article{holme_temporal_2012,
	title = {Temporal networks},
	volume = {519},
	issn = {0370-1573},
	url = {https://www.sciencedirect.com/science/article/pii/S0370157312000841},
	doi = {https://doi.org/10.1016/j.physrep.2012.03.001},
	number = {3},
	journal = {Physics Reports},
	author = {Holme, Petter and Saramäki, Jari},
	year = {2012},
	pages = {97--125},
}

@article{sulem_graph_2024,
	title = {Graph similarity learning for change-point detection in dynamic networks},
	volume = {113},
	issn = {1573-0565},
	url = {https://doi.org/10.1007/s10994-023-06405-x},
	doi = {10.1007/s10994-023-06405-x},
	language = {en},
	number = {1},
	journal = {Machine Learning},
	author = {Sulem, Déborah and Kenlay, Henry and Cucuringu, Mihai and Dong, Xiaowen},
	month = jan,
	year = {2024},
	pages = {1--44},
}

@article{rossetti_community_2018,
	title = {Community {Discovery} in {Dynamic} {Networks}: {A} {Survey}},
	volume = {51},
	issn = {0360-0300},
	shorttitle = {Community {Discovery} in {Dynamic} {Networks}},
	url = {https://dl.acm.org/doi/10.1145/3172867},
	doi = {10.1145/3172867},
	number = {2},
	journal = {ACM Comput. Surv.},
	author = {Rossetti, Giulio and Cazabet, Rémy},
	month = feb,
	year = {2018},
	pages = {35:1--35:37},
}

@incollection{cazabet_challenges_2023,
	address = {Cham},
	title = {Challenges in {Community} {Discovery} on {Temporal} {Networks}},
	isbn = {978-3-031-30399-9},
	url = {https://doi.org/10.1007/978-3-031-30399-9_10},
	doi = {10.1007/978-3-031-30399-9_10},
	language = {en},
	booktitle = {Temporal {Network} {Theory}},
	publisher = {Springer International Publishing},
	author = {Cazabet, Remy and Rossetti, Giulio},
	editor = {Holme, Petter and Saramäki, Jari},
	year = {2023},
	pages = {185--202},
}

@article{cazabet_evaluating_2021,
	title = {Evaluating community detection algorithms for progressively evolving graphs},
	volume = {8},
	issn = {2051-1329},
	url = {https://doi.org/10.1093/comnet/cnaa027},
	doi = {10.1093/comnet/cnaa027},
	number = {6},
	journal = {Journal of Complex Networks},
	author = {Cazabet, Remy and Boudebza, Souâad and Rossetti, Giulio},
	month = mar,
	year = {2021},
	pages = {cnaa027},
}

@article{perra_activity_2012,
	title = {Activity driven modeling of time varying networks},
	volume = {2},
	copyright = {2012 The Author(s)},
	issn = {2045-2322},
	url = {https://www.nature.com/articles/srep00469},
	doi = {10.1038/srep00469},
	language = {en},
	number = {1},
	journal = {Scientific Reports},
	publisher = {Nature Publishing Group},
	author = {Perra, N. and Gonçalves, B. and Pastor-Satorras, R. and Vespignani, A.},
	month = jun,
	year = {2012},
	pages = {469},
}

@article{bhattacharjee_change_2020,
	title = {Change {Point} {Estimation} in a {Dynamic} {Stochastic} {Block} {Model}},
	volume = {21},
	issn = {1533-7928},
	url = {http://jmlr.org/papers/v21/18-814.html},
	number = {107},
	journal = {Journal of Machine Learning Research},
	author = {Bhattacharjee, Monika and Banerjee, Moulinath and Michailidis, George},
	year = {2020},
	pages = {1--59},
}

@book{levin_markov_2017,
	title = {Markov chains and mixing times},
	volume = {107},
	isbn = {1-4704-2962-4},
	publisher = {American Mathematical Soc.},
	author = {Levin, David A. and Peres, Yuval},
	year = {2017},
}

@article{latapy_stream_2018,
	title = {Stream graphs and link streams for the modeling of interactions over time},
	volume = {8},
	issn = {1869-5469},
	url = {https://doi.org/10.1007/s13278-018-0537-7},
	doi = {10.1007/s13278-018-0537-7},
	language = {en},
	number = {1},
	journal = {Social Network Analysis and Mining},
	author = {Latapy, Matthieu and Viard, Tiphaine and Magnien, Clémence},
	month = oct,
	year = {2018},
	pages = {61},
}

@article{gauvin_randomized_2022,
	title = {Randomized {Reference} {Models} for {Temporal} {Networks}},
	volume = {64},
	issn = {0036-1445},
	url = {https://epubs.siam.org/doi/10.1137/19M1242252},
	doi = {10.1137/19M1242252},
	number = {4},
	journal = {SIAM Review},
	publisher = {Society for Industrial and Applied Mathematics},
	author = {Gauvin, Laetitia and Génois, Mathieu and Karsai, Márton and Kivelä, Mikko and Takaguchi, Taro and Valdano, Eugenio and Vestergaard, Christian L.},
	month = nov,
	year = {2022},
	pages = {763--830},
}

@article{snijders_introduction_2010,
	series = {Dynamics of {Social} {Networks}},
	title = {Introduction to stochastic actor-based models for network dynamics},
	volume = {32},
	issn = {0378-8733},
	url = {https://www.sciencedirect.com/science/article/pii/S0378873309000069},
	doi = {10.1016/j.socnet.2009.02.004},
	number = {1},
	journal = {Social Networks},
	author = {Snijders, Tom A. B. and van de Bunt, Gerhard G. and Steglich, Christian E. G.},
	month = jan,
	year = {2010},
	pages = {44--60},
}

@article{eick_physics_1993,
	title = {THE PHYSICS OF THE ${M}_t/{G}/\infty$ QUEUE},
	volume = {41},
	number = {4},
	journal = {Operations Research},
	author = {Eick, Stephen G. and Massey, William A. and Whitt, Ward},
	year = {1993},
	pages = {731},
}

@book{gross_fundamentals_2008,
	title = {Fundamentals of queueing theory},
	isbn = {81-265-1777-8},
	publisher = {John wiley \& sons},
	author = {Gross, Donald},
	year = {2008},
}

@article{dowson_maximum-entropy_1973,
	title = {Maximum-entropy distributions having prescribed first and second moments ({Corresp}.)},
	volume = {19},
	issn = {1557-9654},
	url = {https://ieeexplore.ieee.org/document/1055060},
	doi = {10.1109/TIT.1973.1055060},
	number = {5},
	journal = {IEEE Transactions on Information Theory},
	author = {Dowson, D. and Wragg, A.},
	month = sep,
	year = {1973},
	pages = {689--693},
}

@article{failla_describing_2024,
	title = {Describing group evolution in temporal data using multi-faceted events},
	volume = {113},
	issn = {1573-0565},
	url = {https://doi.org/10.1007/s10994-024-06600-4},
	doi = {10.1007/s10994-024-06600-4},
	language = {en},
	number = {10},
	journal = {Machine Learning},
	author = {Failla, Andrea and Cazabet, Rémy and Rossetti, Giulio and Citraro, Salvatore},
	month = oct,
	year = {2024},
	pages = {7591--7615},
}

@article{kendall_stochastic_1953,
	title = {Stochastic {Processes} {Occurring} in the {Theory} of {Queues} and their {Analysis} by the {Method} of the {Imbedded} {Markov} {Chain}},
	volume = {24},
	issn = {0003-4851, 2168-8990},
	url = {https://projecteuclid.org/journals/annals-of-mathematical-statistics/volume-24/issue-3/Stochastic-Processes-Occurring-in-the-Theory-of-Queues-and-their/10.1214/aoms/1177728975.full},
	doi = {10.1214/aoms/1177728975},
	number = {3},
	journal = {The Annals of Mathematical Statistics},
	publisher = {Institute of Mathematical Statistics},
	author = {Kendall, David G.},
	month = sep,
	year = {1953},
	pages = {338--354},
}

@article{bunday_gmr_1980,
	title = {The {G}/{M}/r machine interference model},
	volume = {4},
	number = {6},
	journal = {European Journal of Operational Research},
	publisher = {Elsevier},
	author = {Bunday, B. D. and Scraton, R. E.},
	year = {1980},
	note = {ISBN: 0377-2217},
	pages = {399--402},
}

@article{little_proof_1961,
	title = {A proof for the queuing formula: ${L} = \lambda {W}$},
	volume = {9},
	number = {3},
	journal = {Operations research},
	publisher = {INFORMS},
	author = {Little, John DC},
	year = {1961},
	note = {ISBN: 0030-364X},
	pages = {383--387},
}

@inproceedings{asgari_mosaic_2023,
	title = {Mosaic benchmark networks: {Modular} link streams for testing dynamic community detection algorithms},
	booktitle = {International {Conference} on {Complex} {Networks} and {Their} {Applications}},
	publisher = {Springer},
	author = {Asgari, Yasaman and Cazabet, Rémy and Borgnat, Pierre},
	year = {2023},
	pages = {209--222},
}

@phdthesis{rossetti_social_2015,
	title = {Social network dynamics},
	url = {https://tesidottorato.depositolegale.it/handle/20.500.14242/149977},
	school = {Università degli Studi di Pisa},
	author = {Rossetti, Giulio},
	year = {2015},
}

@article{casteigts_time-varying_2012,
	title = {Time-varying graphs and dynamic networks},
	volume = {27},
	number = {5},
	journal = {International Journal of Parallel, Emergent and Distributed Systems},
	publisher = {Taylor \& Francis},
	author = {Casteigts, Arnaud and Flocchini, Paola and Quattrociocchi, Walter and Santoro, Nicola},
	year = {2012},
	note = {ISBN: 1744-5760},
	pages = {387--408},
}

@unpublished{barucca_maximum_2025,
	title = {Maximum entropy temporal networks},
	publisher = {arXiv},
	author = {Barucca, Paolo},
	month = oct,
	year = {2025},
    note = {https://doi.org/10.48550/arXiv.2509.02098},
}

@book{walrand_introduction_1988,
	title = {An {Introduction} to {Queueing} {Networks}},
	isbn = {978-0-13-474487-2},
	url = {https://books.google.ch/books?id=CGBRAAAAMAAJ},
	publisher = {Prentice Hall},
	author = {Walrand, J.},
	year = {1988},
}

@book{robert_stochastic_2013,
	title = {Stochastic {Networks} and {Queues}},
	isbn = {978-3-662-13052-0},
	url = {https://books.google.ch/books?id=MCL4CAAAQBAJ},
	publisher = {Springer Berlin Heidelberg},
	author = {Robert, P.},
	year = {2013},
}

@book{asmussen_applied_2003,
	title = {Applied {Probability} and {Queues}},
	isbn = {978-0-387-00211-8},
	url = {https://books.google.ch/books?id=BeYaTxesKy0C},
	publisher = {Springer},
	author = {Asmussen, S.},
	year = {2003},
}

@unpublished{koovely_evolution_2025,
	title = {Evolution of {Conditional} {Entropy} for {Diffusion} {Dynamics} on {Graphs}},
	publisher = {arXiv},
	author = {Koovely, Samuel and Bovet, Alexandre},
	year = {2025},
    note = {https://doi.org/10.48550/arXiv.2510.19441}
}

@article{clemente_temporal_2024,
	title = {Temporal networks with node-specific memory: {Unbiased} inference of transition probabilities, relaxation times, and structural breaks},
	volume = {6},
	issn = {2643-1564},
	shorttitle = {Temporal networks with node-specific memory},
	url = {https://link.aps.org/doi/10.1103/PhysRevResearch.6.043257},
	doi = {10.1103/PhysRevResearch.6.043257},
	language = {en},
	number = {4},
	journal = {Physical Review Research},
	author = {Clemente, Giulio Virginio and Tessone, Claudio J. and Garlaschelli, Diego},
	month = dec,
	year = {2024},
	pages = {043257},
}

@book{masuda_guide_2016,
	title = {A {Guide} {To} {Temporal} {Networks}},
	isbn = {978-1-78634-116-7},
	url = {https://books.google.ch/books?id=84n4DAAAQBAJ},
	publisher = {World Scientific Publishing Company},
	author = {Masuda, N. and Lambiotte, R.},
	year = {2016},
}

@article{presigny_building_2021,
	title = {Building surrogate temporal network data from observed backbones},
	volume = {103},
	number = {5},
	journal = {Physical Review E},
	publisher = {APS},
	author = {Presigny, Charley and Holme, Petter and Barrat, Alain},
	year = {2021},
	note = {ISBN: 2470-0045},
	pages = {052304},
}

@article{longa_efficient_2022,
	title = {An efficient procedure for mining egocentric temporal motifs},
	volume = {36},
	issn = {1573-756X},
	url = {https://doi.org/10.1007/s10618-021-00803-2},
	doi = {10.1007/s10618-021-00803-2},
	number = {1},
	journal = {Data Mining and Knowledge Discovery},
	author = {Longa, Antonio and Cencetti, Giulia and Lepri, Bruno and Passerini, Andrea},
	month = jan,
	year = {2022},
	pages = {355--378},
}

@article{sanna_passino_modelling_2019,
	title = {Modelling dynamic network evolution as a {Pitman}-{Yor} process},
	volume = {1},
	issn = {2639-8001},
	url = {http://aimsciences.org//article/doi/10.3934/fods.2019013},
	doi = {10.3934/fods.2019013},
	language = {en},
	number = {3},
	urldate = {2026-02-10},
	journal = {Foundations of Data Science},
	author = {Sanna Passino, Francesco and Heard, Nicholas A. and {,Department of Mathematics, Imperial College London, 180 Queen's Gate, London – SW7 2AZ, United Kingdom}},
	year = {2019},
	pages = {293--306},
}

@article{sanna_passino_mutually_2023,
	title = {Mutually {Exciting} {Point} {Process} {Graphs} for {Modeling} {Dynamic} {Networks}},
	volume = {32},
	issn = {1061-8600},
	url = {https://doi.org/10.1080/10618600.2022.2096048},
	doi = {10.1080/10618600.2022.2096048},
	number = {1},
	urldate = {2026-02-10},
	journal = {Journal of Computational and Graphical Statistics},
	publisher = {Taylor \& Francis},
	author = {Sanna Passino, Francesco and Heard, Nicholas A.},
	month = jan,
	year = {2023},
	pages = {116--130},
}

\end{document}